\documentclass[12pt]{iopart}
\usepackage{iopams}
\usepackage{graphicx}
\usepackage{subeqn}

\usepackage{subfigure}
\def\be{\begin{equation}}
\def\ee{\end{equation}}
\def\e#1{\label{#1}\end{equation}}
\def\bea{\begin{eqnarray}}
\def\eea{\end{eqnarray}}
\def\ea#1{\label{#1}\end{eqnarray}}

\def\bem#1{\begin{mathletters}\label{#1}}
\def\eml{\end{mathletters}}

\def\ket#1{{|#1\rangle}}

\def\4#1{{\boldsymbol{#1}}}
\def\8#1{{\widetilde{#1}}}

\newcommand{\eqref}{\ref}

\begin{document}

\title{Operational path-phase complementarity in single-photon interferometry}

\author{Noam Erez\footnote{nerez@weizmann.ac.il},
Daniel Jacobs\footnote{danielj@weizmann.ac.il},
Gershon Kurizki\footnote{gershon.kurizki@weizmann.ac.il} }
\address{Department of Chemical Physics,
Weizmann Institute of Science, 76100 Rehovot, Israel}

\begin{abstract}
We examine two setups that reveal different operational implications of path-phase complementarity for single photons in a Mach-Zehnder interferometer (MZI). In both setups, the which-way (WW) information is recorded in the polarization state of the photon serving as a ``flying which-way detector''. In the ``predictive'' variant, using a \emph{fixed} initial state, one obtains duality relation between the probability to correctly predict the outcome of either a which-way (WW) or which-phase (WP) measurement (equivalent to the conventional path-distinguishibility-visibility). In this setup, only one or the other (WW or WP) prediction has operational meaning in a single experiment. In the second, ``retrodictive'' protocol, the initial state is secretly selected {\em for each photon} by one party, Alice, among a set of initial states which may differ in the amplitudes and phases of the photon in each arm of the MZI. The goal of the other party, Bob, is to retrodict the initial state by measurements on the photon. Here, a similar duality relation between WP and WW probabilities, governs their {\em simultaneous} guesses in {\em each experimental run}.
\end{abstract}
\maketitle

\section{Introduction}
\label{sec-intro}

The operational content of quantum complementarity is that the uncertainties in the outcomes of measurements corresponding to complementary (non-commuting) observables (on identically prepared ensembles) necessarily have a tradeoff, i.e., satisfy an uncertainty relation. If one tries to measure two complementary observables on a {\em single} system, the measurement of one observable ``disturbs'' the complementary one,i.e., introduces uncertainty in it. 

\begin{figure}[htb]
    \centering
    \subfigure[Phase measurement]{
        \includegraphics[width=.42\columnwidth]{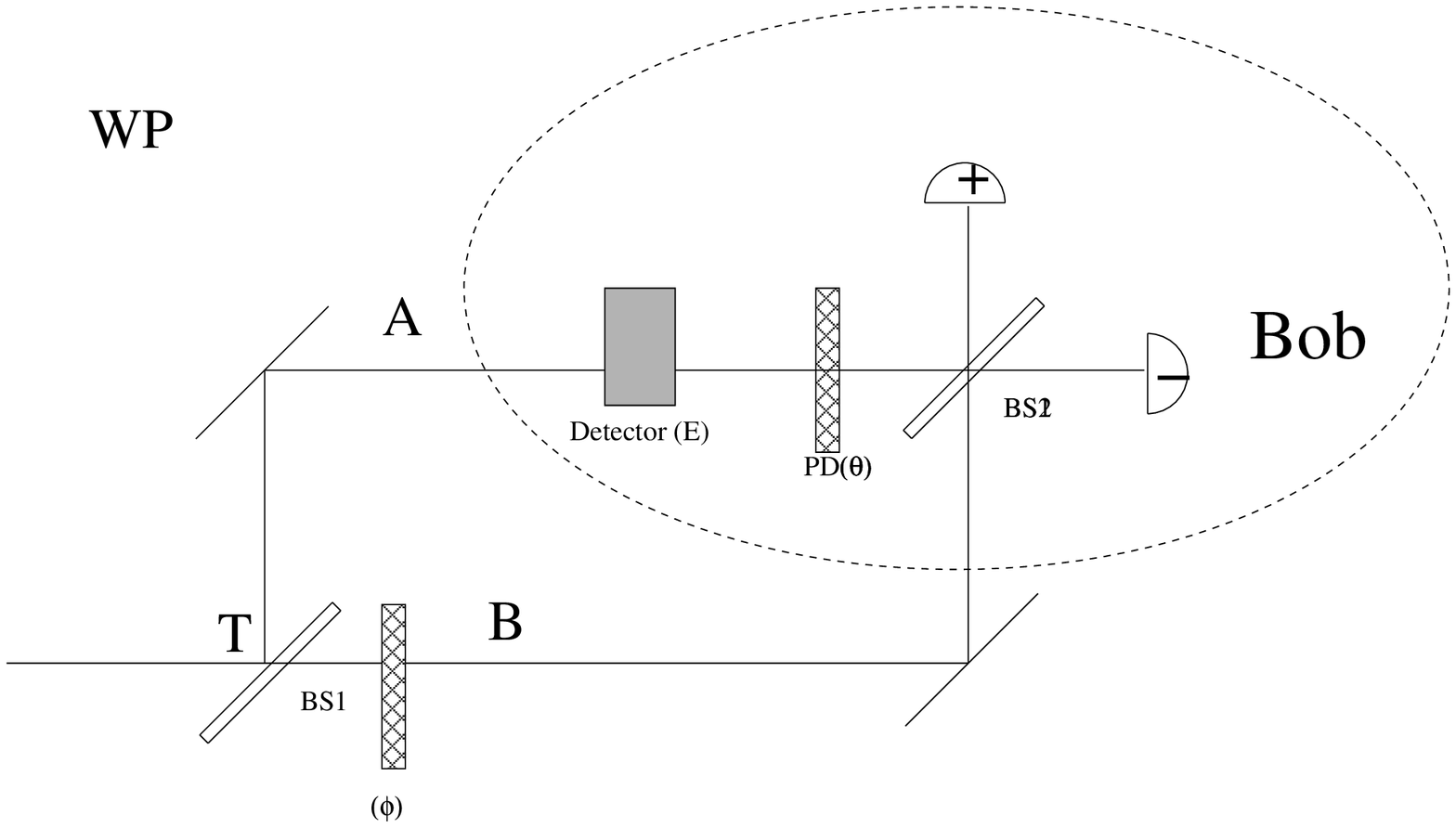}
        \label{ESa}
        }
    \subfigure[Path measurement]{
        \includegraphics[width=.42\columnwidth]{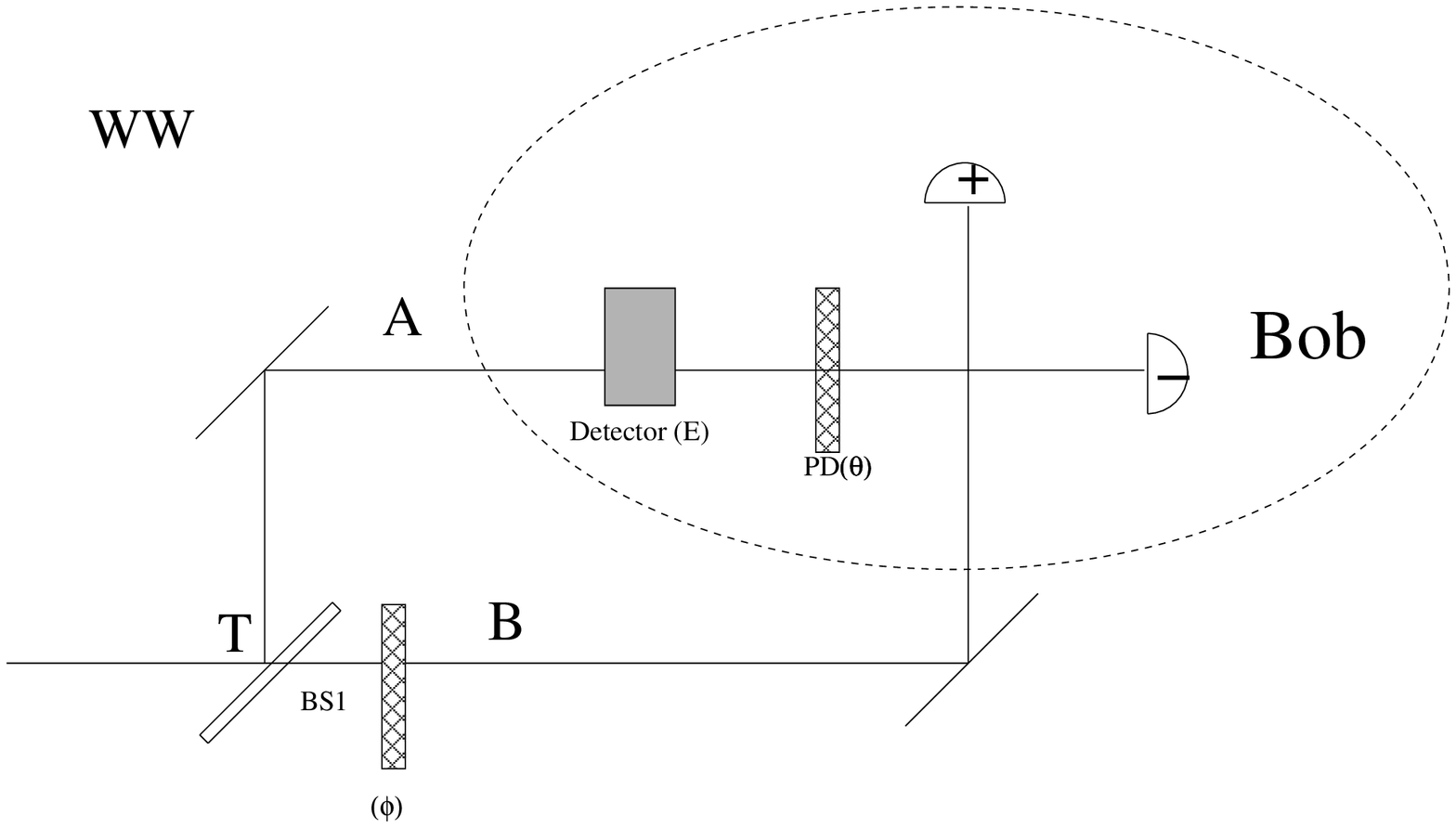}
				\label{ESb} 
				}   
	\caption{MZI with WW-detector. (a) Phase measurement in the presence of an inefficient WW detector. (b) Setup for path measurement.}
	\label{EnglertScheme}
\end{figure}

For example, in the simplest system, with Hilbert space of dimensionality 2, the measurement of one Pauli operator,
say $\sigma_z$, the which-slit observable, ``disturbs'' the ability of the two beams to interfere at a relative phase as
measured by a combination of $\sigma_x$ and $\sigma_y$. 
The quantification of such disturbances has led to the complementarity or duality relation\cite{wheeler1982qta,WZ, GY, JSV, EnglertBG,durr1998fva,jacques:220402} for path predictability versus fringe visibility of a particle in a balanced Mach-Zehnder interferometer (MZI) with a partly efficient which-way detector (Fig. \ref{ESa}). 
This relation reads:
\begin{equation}
\label{eq:std_comp}
D^{2}+V^{2}\leq1
\end{equation}
Here the path distinguishability, $D$, is related to the which-way (WW) probability,
$\mathcal{P}_{WW}$, of guessing the path correctly for a known input state and a WW detector of efficiency (reliability) $E\leq1$. The fringe visibility, $V$, is related to the which-phase (WP) probability, $\mathfrak{\mathcal{P}}_{WP}$, of guessing correctly which MZI port the particle will exit through (for an optimal choice of the phase between the arms) \cite{BergEng, TIE:NJP}:
\begin{equation}
\mathfrak{\mathcal{P}}_{WW}=\frac{1+D}{2},~~\mathfrak{\mathcal{P}}_{WP}=\frac{1+V}{2}.
\label{eq:2}
\end{equation}

Yet, in this setup, the $WW$ and $WP$ probabilities refer to two \emph{alternative} measurements\cite{Luis}. 
Indeed, $\mathcal{P}_{WP}$ is our probability of predicting correctly where the particle will exit (Fig. \ref{ESa}). By contrast, $\mathfrak{\mathcal{P}}_{WW}$ is operationally meaningful only in a measurement (Fig. \ref{ESb}) where the exit beam splitter of the MZI is removed, because only then can the readout of the partly efficient WW detector be
verified.  Thus, in the scheme of Fig. \ref{EnglertScheme} our simultaneous guesses of path and phase cannot be verified or falsified in the same predictive experiment. Rather, the duality relation in Eqs. (\ref{eq:std_comp}) (\ref{eq:2}) describes a trade-off between the predictabilities of two \emph{different} experiments. 

In the WP-experiment, we may think of the WW-detector as {\em counterfactually} ``predicting'' 
what would have occurred in a WW-experiment, had it been performed. However, counterfactual
reasoning is notoriously problematic, as commented by Asher Peres: ``unperformed measurements
have no results''. 

Is it possible to obtain a duality relation for path and phase information which has
a simultaneous operational meaning for each experimental run? Such a relation can indeed be given 
in the context of quantum {\em state discrimination}, namely, measurements aimed at optimally guessing the initial state out of a set of possible states\cite{PhysRevA.65.050305,HerzBerg}. 

To this end, we formulate path-phase complementarity for state-discrimination: the guessing by Bob which of the
alternative input states had been prepared by Alice prior to the measurements Bob performed in a single experimental run. The WW and WP information is then retrodiction by Bob concerning Alice's alternative input states, rather than the standard predictions of possible outcomes of alternative measurements. Both retrodictive and predictive protocols may be implemented using single photons in an MZI, the photons themselves carrying the WW and WP information.


The paper is structured as follows. In Sec. \ref{Sec-2} we describe the proposed setup and the standard predictive protocol followed by Bob and Alice in this setup. Sec. \ref{Sec-3} and \ref{Sec-4} describe two possible retrodictive protocols (using the same experimental setup), and Sec. \ref{Sec-conc} is devoted to conclusions. 

\section{Predictive duality in MZI}
\label{Sec-2}

\subsection{Predictive duality relations for a particle in an MZI}
In the standard formulation of the duality relation, Alice prepares the particle in a known initial state:
\be
|{\rm in}\rangle = \sqrt{w_1} |A\rangle + e^{-i\phi_0}\sqrt{w_2}|B\rangle, 
\ee
where $|A\rangle~(|B\rangle)$ denotes the state of being in arm $A(B)$ of the MZI. It is preferable to use a balanced MZI to maximize the coherence.
The (variable) phase delay element in arm $B$ adds an additional relative phase of $\phi$. 
The output beam splitter effects the transformation:
\be
\left\{ \begin{array}{lll} |A\rangle &\mapsto& \frac{|+\rangle + |-\rangle}{\sqrt{2} } \\ |B\rangle &\mapsto& \frac{|+\rangle - |-\rangle}{\sqrt{2} i}. \end{array} \right.
\ee
where $|\pm\rangle$ are states that correspond to exiting via the two output ports. 
The probability of the (ideal) photon detector in output port $+$ to detect the particle, $P_+$, is a function
of the initial state prepared by Alice and the phase delay $\phi$. If we repeat the experiment many times with
the same initial state but varying $\phi$, we can measure the ``interference pattern'' $P_+(\phi)$. One can then define the visibility of this ``fringe pattern'' as 
\be
V = \frac{\max(P_+)-\min(P_+)}{\max(P_+)+\min(P_+)}.
\ee
In the absence of a WW detector, the only information we have on the probabilities to find the particle in
one of the paths is given by the amplitudes in the initial state, or the corresponding weights $w_1,w_2$. 
If $w_1 \neq w_2$, then there are unequal probabilities of finding the particle in either arm (if such a measurement
is actually performed), while the visibility is reduced. 
The probability of correctly guessing in which arm the particle will be found is given:  
\be
\mathcal{P}_{WW}=\frac{1+P}{2}, 
\ee
where $P$ is the predictability: 
\be
P=\left|w_1 - w_2 \right|.
\ee
The predictability satisfies a duality relation with the visibility\cite{GY}: 
\be
P^2+V^2\leq 1,
\ee
which is just Eq. (\ref{eq:std_comp}) with $D=P$.

The effect of introducing an imperfect WW detector in one or both arms can be modeled as follows\cite{EnglertBG}.
Let the intial (fiducial) state of the detector be denoted by $|0\rangle_D$. Then, after the particle interacts with it, their joint state undergoes the transformation:
\be
\left( \sqrt{w_1} |A\rangle + e^{-i\phi_0}\sqrt{w_2}|B\rangle \right) |0\rangle_D \mapsto 
\sqrt{w_1} |A\rangle |a\rangle_D + e^{-i\phi_0}\sqrt{w_2}|B\rangle |b\rangle_D, 
\ee
where $|a\rangle_D$ and $|b\rangle_D$ are the detector ``pointer''-states, which are {\em not} required to be orthogonal.
More generally, the detector states can be allowed to be mixed ($\rho^0_D,~\rho^a_D,~\rho^b_D$).
Without loss of generality, we shall assume a WW detector to be present in arm $A$ only (as in Fig. (\ref{EnglertScheme}) ).

When there is no a priori bias toward either path ($P=0$), but there is a WW detector, Eq. (\ref{eq:std_comp}) applies with 
\be
D=E=\frac{1}{2}{\rm Tr}\left| \rho^a_D - \rho^b_D \right|.
\label{eq:DE}
\ee
Here $E$ is the detector efficiency\cite{EnglertBG} expressed as the trace norm of the difference of the two detector states (the absolute value of an operator, $O$, is defined as: $|O|\equiv \sqrt{O^\dagger O}$). Equation (\ref{eq:DE}) then gives the probability of correctly guessing the outcome of a WW measurement if we (optimally) measure the state of the detector.
For pure 
detector ``pointer''-states, 
$|a\rangle,~|b\rangle$, this takes the simpler form:
\be 
E= \sqrt{1-|\langle a | b\rangle |^2}.
\label{eq:E}
\ee
In the general case, where the initial state has a priori which-way bias and there is a WW detector, Eq. (\ref{eq:std_comp}) applies, with the path distinguishability, $D$, given by\cite{BergEng}:
\be
D= {\rm Tr} \left| w_1 \rho^a_D - w_2 \rho^b_D \right|.
\ee

For pure ``pointer''-states, $D$ is determined by $P$ and $E$:

\be 
D= \sqrt{P^2 + E^2-E^2 P^2}.
\ee

A few special cases deserve notice: for $E=0$, we have $D=P$; for $P=0$, $D=E$; and for $E=1$ or $P=1$, we have $D=1$.

\subsection{Interferometric setup with TIE detector}

How can a limited-efficiency WW detector be experimentally implemented? 
In their experiments \cite{durr1998fva,Durr}) D{\"u}rr et al. used the internal degrees of freedom of interfering
atoms to store the WW information. Intraparticle translational-internal entanglement (TIE), proposed in \cite{TIE:NJP,TIE:IJPB,bargill2006ssm,TIE:OL} provides a means of simultaneously encoding both WW and WP information
in a ``flying detector'', e.g., a photon in the MZI whose translational (path) and internal (polarization) states are entangled in a four-dimensional space. Recently a single-photon complementarity experiment has been carried out using polarization\cite{jacques:220402}. 

The simplest variant of TIE interferometry involves a particle (photon) whose orthogonal internal states $|1\rangle$ and $|2\rangle$ pick up a phase while traversing an MZI.
In arm $A$ the particle propagates unchanged (up to an overall phase $\beta_0$), whereas in arm $B$, its internal states $|1\rangle,~|2\rangle$ accrue a relative phase proportional to the path length (or in the present scheme, $2\beta$ due to the Faraday rotator):
\be
c_1 |1\rangle + c_2|2\rangle \mapsto e^{i\beta_0}\left( c_1 |1\rangle + e^{2i\beta}c_2 |2\rangle \right). 
\label{eq:TIE}
\ee
The internal states serve as the detector ``pointer''-states. Then Eq. (\ref{eq:E}) gives: 
$E = |c_1 c_2 \sin \beta|$.

As in\cite{TIE:OL}, we here consider a photon as the TIE detector in the MZI, and its polarizations as the detector ``pointer''-states.
\begin{figure}[htb]
\centering\includegraphics[width=0.7\columnwidth]{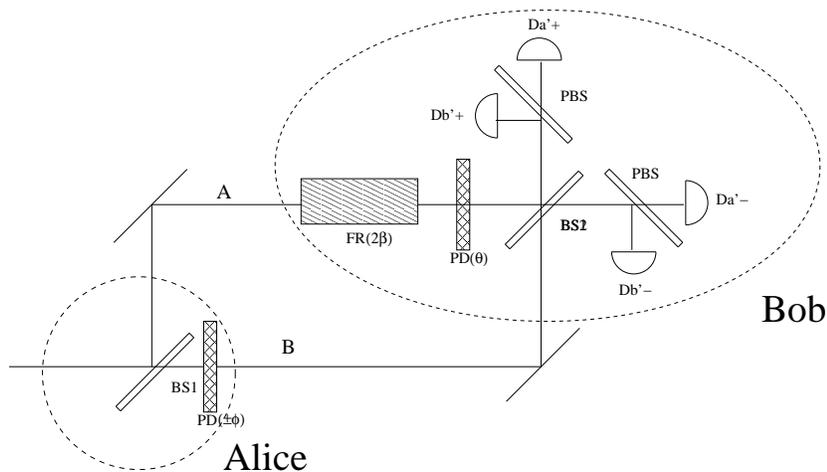}
\caption{Suggested experimental scheme. BS1, BS2 = beam splitters (for retrodictive experiment, BS1 should
be a variable beam splitter); FR = Faraday rotator; PD = (variable) phase-delay; PBS = polarizing beam splitters;
$Da'\pm,~ Db'\pm$ = photon detectors. }
\label{fig:Exp}
\end{figure}
Specifically, we could place a Faraday rotator and phase delay in arm $A$ of the MZI (Fig. (\ref{fig:Exp}) ). The Faraday rotator rotates the polarization plane through the angle $\beta$, while the phase delay is $\theta$.
This would conform to the general TIE scheme in Eq. (\ref{eq:TIE}) with $|1 (2)\rangle$ states corresponding to the right (left) {\em circular} polarization states ($|R (L)\rangle $). For a {\em linearly} polarized input state ($|c_1|=|c_2|=\frac{1}{\sqrt{2} }$), this gives simply 
\be
E = |\sin \beta|.
\ee  

\section{Retrodictive (state-discrimination) duality in MZI}
\label{Sec-3}

\subsection{Protocol}
Alice randomly chooses, using the setup in Fig. (\ref{fig:Exp}), to prepare the qubit in one of the four input states: 

\begin{equation}
\label{eq:inputs}
\ket{b_{ww},b_{wp}}_{\alpha,\phi}\equiv T(b_{ww}\alpha)\ket{A}+e^{b_{wp}i\phi}T(-b_{ww}\alpha)\ket{B}
\end{equation}
Here
\be
T(\pm{}b_{ww}\alpha)=\cos\left(\frac{\pi}{4}\pm\frac{b_{ww}\alpha}{2}\right)
\label{eq:fixed_in}
\ee
are the input amplitudes of $\ket{A}$ and $\ket{B}$, $e^{b_{wp}i\phi}$ their
relative phase factor; both parameterized by
\be
b_{ww}=\pm1,b_{wp}=\pm1. 
\ee

\begin{figure}[htb]
   \centering \includegraphics[width=8cm]{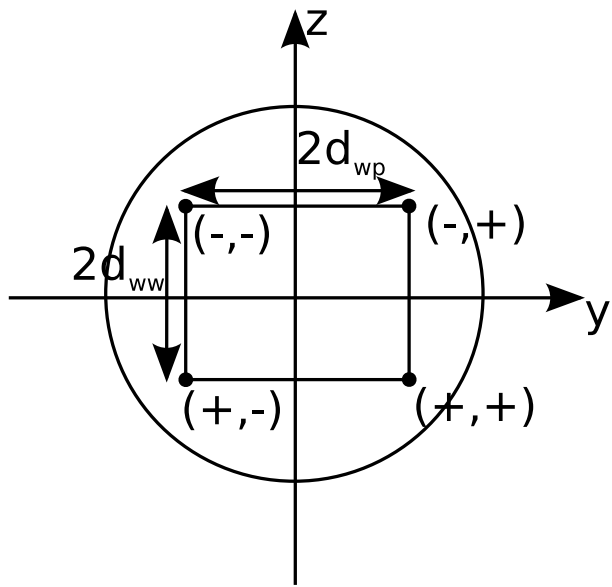}
    \caption{Four alternative input states, $\ket{b_{ww}, b_{wp}}$, plotted on the Bloch Sphere, labeled by the $+,-$ signs of $b_{ww}$ and $b_{wp}$.  The
the geometrical meaning of the trace distances $d_{ww}$ and $d_{wp}$ is shown.}
\label{fig:rect_states}
\end{figure}

In an MZI, ${b_{ww}\alpha}$ affects the bias for propagating along arm A versus
arm B, i.e. having the photon in path state $\ket{A}$ or $\ket{B}$. This bias determines the which-way (WW)
probability.  The parameter $b_{wp}\phi$ affects the relative phase of these
states, i.e., the which-phase (WP) probability.

Figure \ref{fig:rect_states} shows the Bloch representation of the set of input states with the identification of the path states $\ket{A,B}$ as qubit states $\ket{\sigma_{z}=\pm1}$.
Bob receives the qubit, and after performing a measurement of his choice, tries
to guess the values of the two bits $b_{ww},b_{wp}$(which are statistically
independent), i.e., guess which of the four possible input states was chosen by
Alice. 

The two bits specifying Alice's choice of preparation are classical
variables. Similarly, the outcome of any measurements Bob chooses to make are given by classical variables. 
They are all operationally meaningful for each experimental run, and the task of guessing the initial state given the measurement results is a well defined statistical problem. On the other hand, their {\em joint probability distribution} is determined by quantum mechanics, of course.
Since Bob would like to maximize the information provided by his measurements on the initial state, 
he faces a problem of maximizing the classical information content of quantum measurements,similar to the concept of the classical information content of a quantum channel\cite{holevo2001ssq}. 

We look for the set of WW and WP probability pairs $(P_{WW}, P_{WP})$ that are {\em optimal}, in the sense that it is not possible to improve one of the probabilities (e.g., by changing the detector efficiency) without diminishing the other. 
We distinguish these {\em ``retrodictive''} probabilities from the ``predictive'' $(\mathcal{P}_{WW},\mathcal{P}_{WP})$ introduced in Eq. (\ref{eq:2}) above. 


We now prove that for WW detector with efficiency $E$, the complementary retrodictive probabilities are related by 

\begin{subequations}
\label{eqs:ParFront}

\begin{equation}
\label{eq:Pww_eng}
\frac{2P_{WW}-1}{d_{WW}}=E,
\frac{2P_{WP}-1}{d_{WP}}=\sqrt{1-E^{2}}.
\end{equation}
where $d_{WW}$ and $d_{WP}$ are the trace distances between the appropriate input states in Fig.\ref{fig:rect_states}, constrained by
\be
d_{WW}^2+d_{WP}^2 \leq 1.
\label{eq:Pyth}
\ee

\end{subequations}


\subsection{Duality proof}

In the TIE scheme of Fig.\ref{fig:Exp}, Alice's input state is, in the notation of Eq. (\ref{eq:fixed_in}),

\begin{equation}
 \ket{b_{ww},b_{wp}}_{\alpha,\phi}\equiv \left( T(b_{ww}\alpha)\ket{A}+e^{b_{wp}i\phi}T(-b_{ww}\alpha)\ket{B} \right) |b\rangle_{\rm pol}
\end{equation}
where $|b\rangle$ is the initial polarization state, and as before, $|A(B)\rangle$ are the states corresponding to the particle being in arm A(respectively B) of the MZI.

After interacting with the Faraday rotator, it maps to:

\begin{equation}
T(b_{ww}\alpha)\ket{A}|a\rangle_{\rm pol}+ e^{b_{wp}i\phi}T(-b_{ww}\alpha)\ket{B}|b\rangle_{\rm pol},
\end{equation}
where $|a\rangle_{\rm pol}$ and $|b\rangle_{\rm pol}$ are the polarization states correlated to the path states $|A\rangle$, $|B\rangle$, respectively.

\begin{figure}[htb]
   \centering \includegraphics[width=8cm]{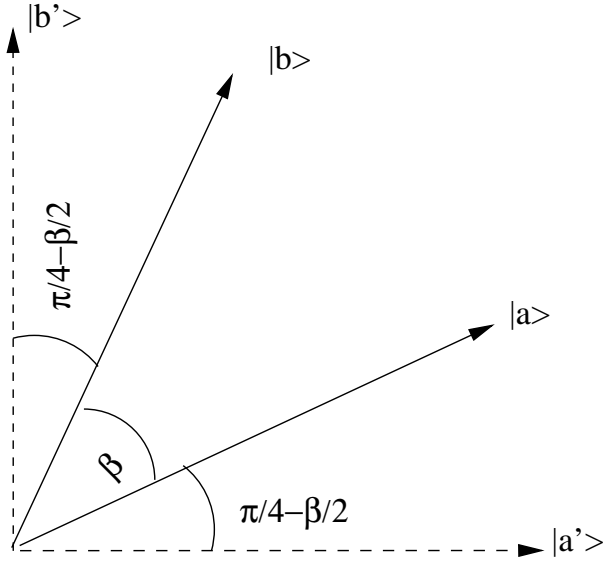}
   \caption{Schematic representation of optimal measurement basis for minimal-error discrimination between two non-orthogonal states $|a\rangle,~~|b\rangle$. When optimally distinguishing between two states there is no advantage to using generalized measurements (POVMs). The orthogonal states $|a'\rangle,~~|b'\rangle$ represent the (orthogonal) projection basis in the optimal measurement\cite{helstrom1969qda}.}
\label{fig:Helstrom}
\end{figure}

The correspondence between states at the output ports of the second beam splitter (BS2 in Fig. \ref{fig:Exp}) and the input states is:
\be
|\pm \rangle_{\rm port} = \frac{|A\rangle \pm i|B\rangle}{\sqrt{2}}
\ee
The measurement basis ${|a'\rangle,~|b'\rangle}$ that optimizes the distinguishability of the detector ``pointer''-states is shown graphically in Fig. \ref{fig:Helstrom}. 

It satisfies the following useful relations:

\begin{subequations}
\bea
 |\langle a' | a \rangle_{\rm pol}|^2 = | \langle b' | b \rangle_{\rm pol}|^2 = \cos^2(\frac{\pi}{4}-\frac{\beta}{2}) 
 = \frac{1+E}{2} \\
 |\langle a' | b \rangle_{\rm pol}|^2 = | \langle b' | a \rangle_{\rm pol}|^2 = \cos^2(\frac{\pi}{4}+\frac{\beta}{2})
 = \frac{1-E}{2}. 
\eea
\label{eq:geometry}
\end{subequations}

Bob's measurement basis is: 
\be
\ket{b_{ww}^o,b_{wp}^o} = \ket{b_{ww}^o}_{\rm port}\ket{b_{wp}^o}_{\rm pol},
\label{eq:meas_bas}
\ee
where 
\begin{subequations}
\label{eqs:24}
\bea
|b_{wp}^{o} &=& +1(-1)\rangle_{\rm pol}\equiv |a'(b')\rangle_{\rm pol};\\
|b_{ww}^{o} &=& +1(-1)\rangle_{\rm pol}\equiv |+1(-1)\rangle_{\rm port}.
\eea
\end{subequations}

These basis states correspond to the 4 output ports in Fig. \ref{fig:Exp}.

The joint input-output probabilities are given by:

\be
P_{b_{ww},b_{wp};b_{ww}^{out},b_{wp}^{out}} = \left|\langle b_{ww}^o,b_{wp}^o | b_{ww},b_{wp} \rangle \right|^2.
\ee

Explicitly (using Eqs. (\ref{eq:geometry})-(\ref{eqs:24}) ):
\be
P_{b_{ww},b_{wp};b_{ww}^{out},b_{wp}^{out}} = \frac{1}{16}
\left(1 + (-1)^{b_{ww}^{out}-b_{ww}}Ed_{WW} +(-1)^{b_{wp}^{out}-b_{wp}}\sqrt{1-E^2}d_{WP} \right)
\ee

The WW and WP marginal probability distributions are:

\be
P_{b_{ww};b_{ww}^{out}} \equiv \sum_{b_{wp},b^{out}_{wp}}P_{b_{ww},b_{wp};b_{ww}^{out},b_{wp}^{out}}=
\frac{1}{4} \left(1 + (-1)^{b_{ww}^{out}-b_{ww}}Ed_{WW} \right)
\ee
and similarly
\be
P_{b_{wp};b_{wp}^{out}} = \frac{1}{4}
\left(1 + (-1)^{b_{wp}^{out}-b_{wp}}\sqrt{1-E^2}d_{WP} \right)
\ee

The probabilities of inferring the WW and WP input bits are:

\be 
P_{WW} = \sum_{b_{ww}}\max_{b_{ww}^{out}}P_{b_{ww};b_{ww}^{out}} = \frac{1}{2} \left(1 + Ed_{WW} \right),
\ee

\be 
P_{WP} =  \frac{1}{2} \left(1 + \sqrt{1-E^2}d_{WP} \right),
\ee
$d_{WW}$ being the trace distance between two input states with the same
value of $b_{WP}$ and different $b_{WW}$, and conversely for $d_{WP}$:
\bea
d_{WW} &\equiv& d_{\rm Trace}(|b_{ww}=+1,b_{wp}\rangle,|b_{ww}=-1,b_{wp}\rangle),  \nonumber \\
d_{WP} &\equiv& d_{\rm Trace}(|b_{ww},b_{wp}=+1\rangle,|b_{ww},b_{wp}=-1\rangle).
\eea
This completes the proof of Eq. (\ref{eq:Pww_eng}) for the set of pairs $(P_{WW},P_{WP})$ that one achieves in the MZI with the WW detector upon varying the detector efficiency, $E$,
We shall show elsewhere that this equality is a bound on all possible measurement schemes, which in general
satisfy the inequality: 
\begin{equation}
\label{eq:ellipse2}
\left(\frac{2P_{WW}-1}{d_{WW}}\right)^{2}+\left(\frac{2P_{WP}-1}{d_{WP}}\right)^{2} \leq 1.
\end{equation}





In particular, the measurement procedure described here (and in Sec. \ref{Sec-2}) defines a {\em generalized measurement} (POVM)\cite{peres1995qtc} on the initial translational (path) state of the photon, in a Hilbert space of dimension 2. A Von-Neumann (VN) measurement of this state would yield 2 possible outcomes, yet the present scheme has 4 possible measurement outcomes. This comes about by first performing a joint unitary operation on the translational state and the polarization state of the photon, and then performing a VN measurement on {\em both}. The effect of the measurement on the translational (path) state alone is then:
\be
\rho_{\rm in} \mapsto \frac{K_i \rho_{\rm in} K_i^\dagger}{Tr \{ K_i \rho_{\rm in} K_i^\dagger \} } i =1,\ldots,4,
\ee
where outcome $i$ occurs with probability $Tr \{ K_i \rho_{\rm in} K_i^\dagger \}$, and the four Kraus operators, $K_i$ are given by:

\begin{subequations}
\bea
K_{a'\pm} = \frac{1}{2}|\pm\rangle \left[ \sqrt{1+E} \langle A| \pm i \sqrt{1-E} \langle B| \right] \\
K_{b'\pm} = \frac{1}{2}|\pm\rangle \left[ \sqrt{1-E} \langle A| \pm i \sqrt{1+E} \langle B| \right].
\eea
\end{subequations}
It is straightforward to verify that the output states are the same as the measurement basis states (Eq. (\ref{eq:meas_bas}) ), and that the operators $A_i \equiv K_i^\dagger K_i$ are positive (non-orthogonal) operators whose sum is the identity, they have the properties of a POVM\cite{peres1995qtc}.

\section{Alternative path or phase retrodiction protocol} 
\label{Sec-4}

The retrodictive guessing game in Sec. \ref{Sec-3} is not, however, simply a time reversal
of the conventional predictive one. To illustrate this, let us consider
yet another ``guessing game'' with alternative input states, which is much
closer to a time-reversed version of the latter. 

\begin{figure}[htb]
    \centering
    \subfigure[WP]{
        \includegraphics[width=.42\columnwidth]{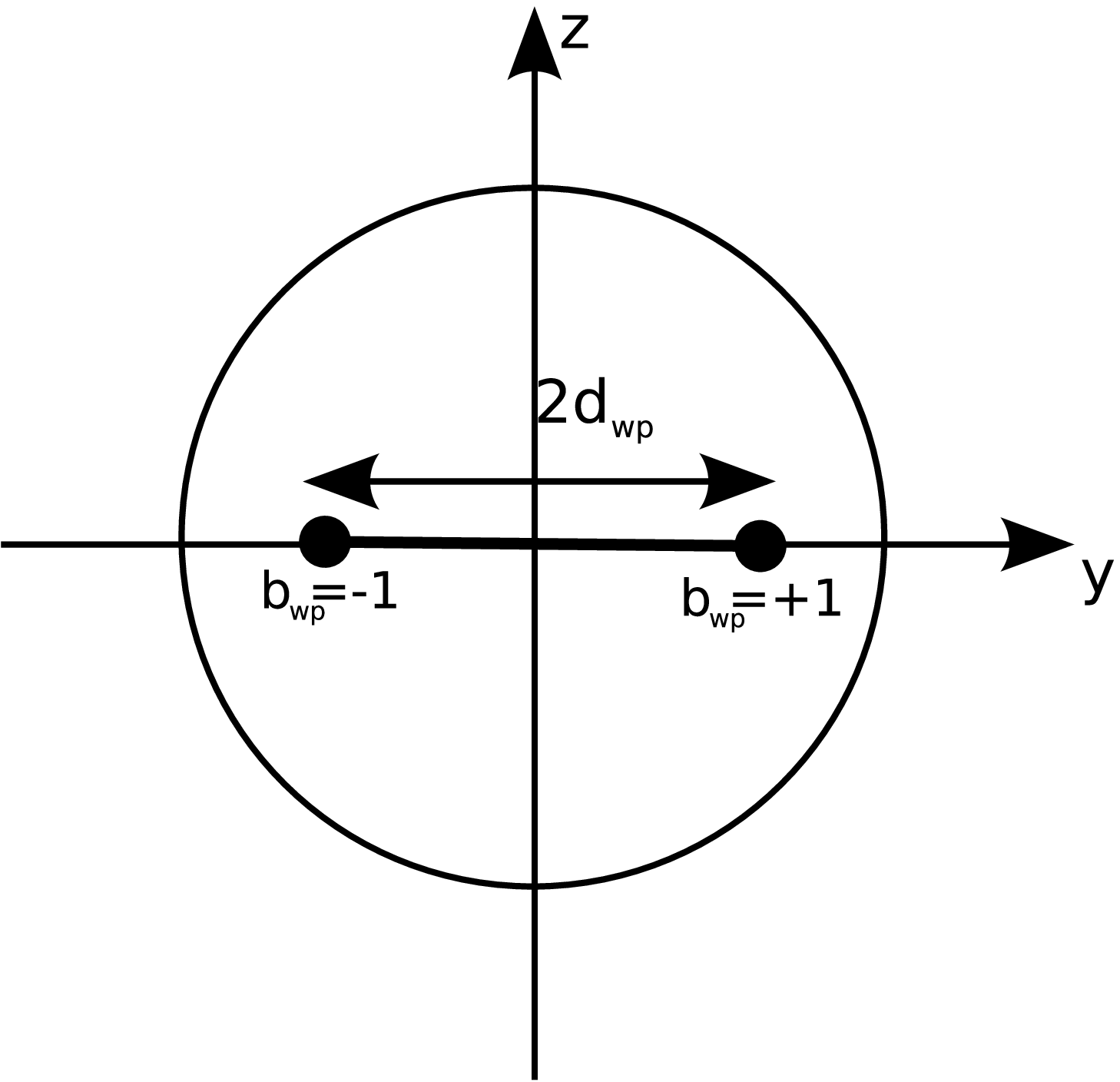}
        \label{fig:alt_int_wp}
        }
    \subfigure[WW]{
        \includegraphics[width=.42\columnwidth]{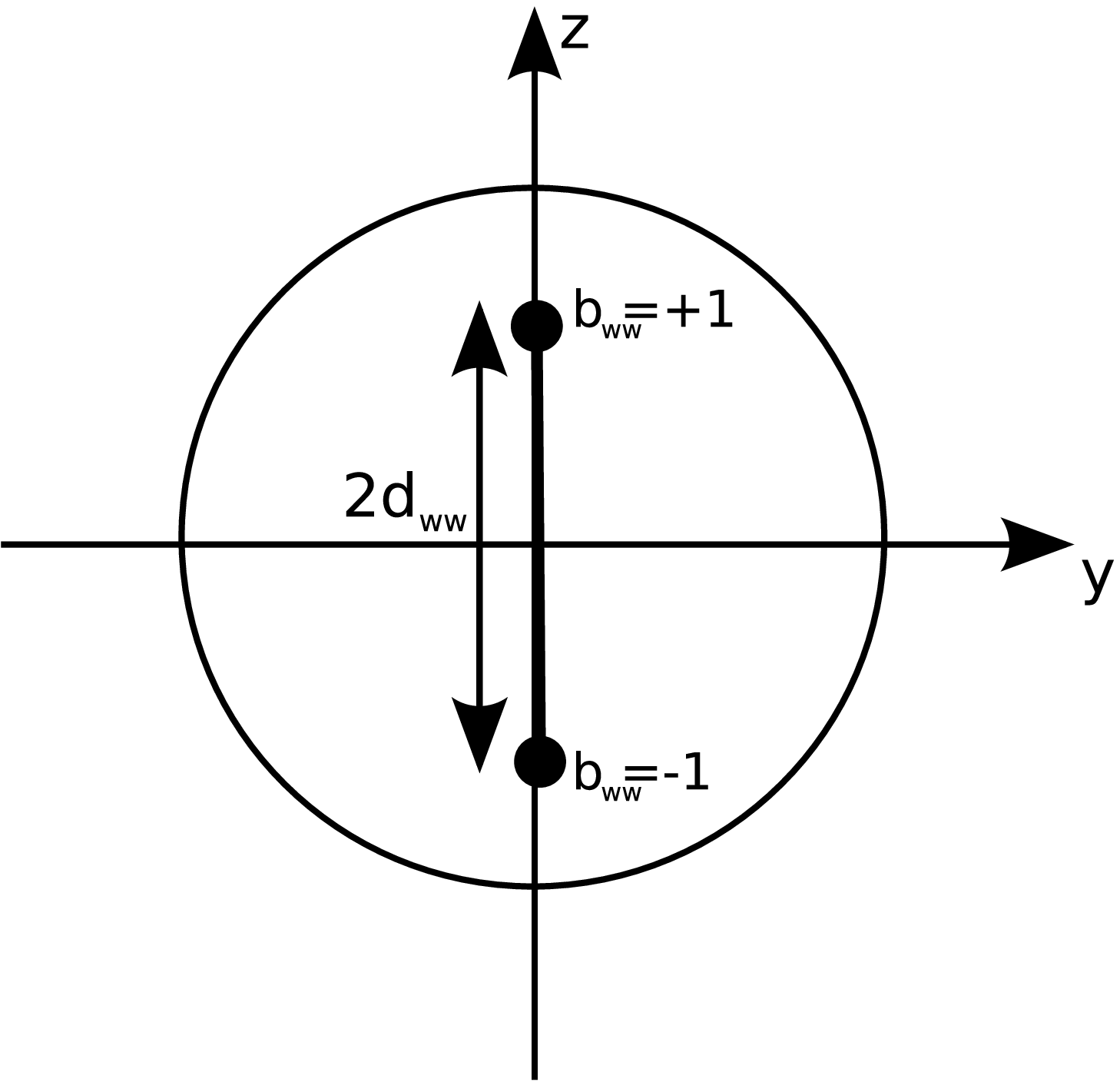}
				\label{fig:alt_int_ww} 
				}   
    \caption{In the conventional, predictive guessing game, Bob must make one
    of two measurements, either he measures the phase \subref{fig:alt_int_wp}
    or the path \subref{fig:alt_int_wp}. This contrasts with the input states of Fig. \ref{fig:rect_states}. }
    \label{fig:alt_ints}
\end{figure}

The predictability-visibility complementarity relation, Eq. (\ref{eq:std_comp}), can be rewritten as:
\begin{equation}
\label{eq:alt_ellipse}
\left(2\mathfrak{\mathcal{P}}_{WW}-1\right)^{2}+\left(2\mathfrak{\mathcal{P}}_{WP}-1\right)^{2}\leq1,
\end{equation}
using Eq. (\ref{eq:2}). This would be the same as our Eq. (\ref{eq:ellipse2}) if we had
$d_{WW}=d_{WP}=1$. This, however, is not possible in our scenario, since it contradicts
the additional constraint (\ref{eq:Pyth}). Hence, our retrodictive
guessing game is not simply a time-reversed version of the conventional
predictive scheme: different sets of states serve for measurements in the
conventional scheme and in the present one (see Fig. \ref{fig:alt_ints}).

Let us, therefore, formulate a game which is closer to a time-reversal of the conventional predictive one, and calculate the path-phase probability constraints for it. 
Like the coventional ``predictive'' guessing game, in which each experimental run involves only one
type of measurement (path or phase), the state preparation now involves either alternative paths, or alternative phases
in each experiment.
Namely, in each run Alice first randomly chooses whether to prepare a which-way (WW) or a which-phase (WP) input state.  This preparation basis is the same as Bob's measurement basis in the \emph{conventional}
scheme (Fig. \ref{fig:alt_ints}). Bob decides on his measurement strategy,
ignorant of Alice's choice. Once Bob has performed his measurement, Alice tells
which basis she has used, and then Bob has to make his guess (as in the BB84
encryption scheme). Alice's two preparation bases correspond to the degenerate
cases: $\left(d_{WW},d_{WP}\right)$ equal to $\left(1,0\right)$ or
$\left(0,1\right)$ for WW or WP, respectively. 

Now let us assume Bob has chosen, as in the conventional case, a certain detector efficiency $E$. Then, for Alice's WW
preparation, Eq. (\ref{eq:Pww_eng}) reduces to
\begin{equation}
\left(2P_{WW}^{\prime}-1\right)^{2}=E^{2}
\end{equation}
 and for Alice's WP preparation to:
\begin{equation}
\left(2P_{WP}^{\prime}-1\right)^{2}=1-E^{2}
\end{equation}
(where the primes are used to distinguish this game from that discussed in Sec.
\ref{Sec-3}). These two equations together imply, of
course,
\begin{equation}
\left(2P_{WW}^{\prime}-1\right)^{2}+\left(2P_{WP}^{\prime}-1\right)^{2}=1,
\end{equation}
which is formally the same as Eq. (\ref{eq:alt_ellipse}) above with equality
sign!
Hence this type of retrodictive guessing game, unlike the one discussed in Sec. \ref{Sec-3}, manifests the same tradeoff between path and phase guess probabilities as Eq. (\ref{eq:std_comp}), derived for the conventional predictive guessing game.

However, for a fair comparison with Eq. (\ref{eq:Pww_eng}), we note that if the
WW and WP bases are degenerate rectangles, we could at each run ask Bob to
guess both path and phase. Then, when Alice has chosen the WW basis his
probability of guessing WW correctly would be the same as above, and would be
equal to $\frac{1}{2}$ when she has chosen the WP basis (and conversely for
Bob's WP guesses). The properly averaged probabilities would then satisfy:
\begin{equation}
\left(2\overline{P_{WW}^{\prime}}-1\right)^{2}+\left(2\overline{P_{WP}^{\prime}}-1\right)^{2}=\left(\frac{1}{2}\right)^{2}
\end{equation}
This is {\em worse} than our bound in Eq. (\ref{eq:ellipse2}), e.g., for the case
where $d_{ww}=d_{wp}=\frac{1}{\sqrt{2}}$.

\section{Conclusions}
\label{Sec-conc}

We have examined two protocols that reveal different operational implications of path-phase complementarity for single photons in a Mach-Zehnder interferometer (MZI). Both protocols use a setup where the which-way (WW) information is recorded in the polarization state of the photon serving as a ``flying which-way detector'', by virtue of its translational-internal entanglement (TIE) \cite{TIE:NJP,TIE:IJPB,bargill2006ssm,TIE:OL}. In the ``predictive'' protocol, using a \emph{fixed} initial state, one obtains duality relation between the probability to correctly predict the outcome of either a which-way (WW) or which-phase (WP) measurement (governed by the conventional duality of path-distinguishibility and visibility). In this setup, only one or the other (WW or WP) prediction has operational meaning in a single experiment. In this protocol, ``retrodictive'' protocol, the initial state is secretly selected {\em for each photon} by one party, Alice, among a set of initial states which may differ in the amplitudes and phases of the photon in each arm of the MZI. The goal of the other party, Bob, is to retrodict the initial state by measurements on the photon. Here, a similar duality relation between WP and WW probabilities, governs their {\em simultaneous} guesses in {\em each experimental run}. 

The restatement of complementarity as a state discrimination problem opens the way to its information-theoretic formulation, to be discussed elsewhere. On the applied side, our guessing games may be the basis of quantum cryptographic schemes. 


\ack We acknowledge the support of the ISF and EC.

\section*{References}
\bibliography{PhotonComp_arx}

\begin{thebibliography}{10}

\bibitem{wheeler1982qta}
J.A. Wheeler and W.H. Zurek.
\newblock {\em {Quantum theory and measurement}}.
\newblock Princeton University Press New Jersey, 1982.

\bibitem{WZ}
William~K. Wootters and Wojciech~H. Zurek.
\newblock Complementarity in the double-slit experiment: Quantum
  nonseparability and a quantitative statement of bohr's principle.
\newblock {\em Phys. Rev. D}, 19(2):473--484, Jan 1979.

\bibitem{GY}
D.~M. {Greenberger} and A.~{Yasin}.
\newblock {Simultaneous wave and particle knowledge in a neutron
  interferometer}.
\newblock {\em Physics Letters A}, 128:391--394, April 1988.

\bibitem{JSV}
Gregg Jaeger, Abner Shimony, and Lev Vaidman.
\newblock Two interferometric complementarities.
\newblock {\em Phys. Rev. A}, 51(1):54--67, Jan 1995.

\bibitem{EnglertBG}
Berthold-Georg Englert.
\newblock Fringe visibility and which-way information: An inequality.
\newblock {\em Phys. Rev. Lett.}, 77(11):2154--2157, Sep 1996.

\bibitem{durr1998fva}
S.~D{\"u}rr, T.~Nonn, and G.~Rempe.
\newblock {Fringe Visibility and Which-Way Information in an Atom
  Interferometer}.
\newblock {\em Physical Review Letters}, 81(26):5705--5709, 1998.

\bibitem{jacques:220402}
Vincent Jacques, E~Wu, Fr\'{e}d\'{e}ric Grosshans, Fran\c{c}ois Treussart,
  Philippe Grangier, Alain Aspect, and Jean-Fran\c{c}ois Roch.
\newblock Delayed-choice test of quantum complementarity with interfering
  single photons.
\newblock {\em Phys. Rev. Lett.}, 100:220402, 2008.

\bibitem{BergEng}
B.-G. {Englert} and J.~A. {Bergou}.
\newblock {Quantitative quantum erasure}.
\newblock {\em Optics Communications}, 179:337--355, May 2000.

\bibitem{TIE:NJP}
Michal Kol\'{a}r, Tom\'{a}s Opatrn\'{y}, Nir Bar-Gill, Noam Erez, and Gershon
  Kurizki.
\newblock Path-phase duality with intraparticle translational-internal
  entanglement.
\newblock {\em New Journal of Physics}, 9(5):129, 2007.

\bibitem{Luis}
Alfredo Luis.
\newblock Operational approach to complementarity and duality relations.
\newblock {\em Phys. Rev. A}, 70(6):062107, Dec 2004.

\bibitem{PhysRevA.65.050305}
Ulrike Herzog and J\'anos~A. Bergou.
\newblock Minimum-error discrimination between subsets of linearly dependent
  quantum states.
\newblock {\em Phys. Rev. A}, 65(5):050305, May 2002.

\bibitem{HerzBerg}
Ulrike Herzog and J\'anos~A. Bergou.
\newblock Distinguishing mixed quantum states: Minimum-error discrimination
  versus optimum unambiguous discrimination.
\newblock {\em Phys. Rev. A}, 70(2):022302, Aug 2004.

\bibitem{Durr}
S.~{D{\"u}rr}, T.~{Nonn}, and G.~{Rempe}.
\newblock Origin of quantum-mechanical complementarity probed by a `which-way'
  experiment in an atom interferometer.
\newblock {\em Nature}, 395:33--37, September 1998.

\bibitem{TIE:IJPB}
Michal Kol\'{a}r, Tom\'{a}s Opatrn\'{y}, Nir Bar-Gill, and Gershon Kurizki.
\newblock Betting on interferometric paths and phases using
  translational-internal entanglement: The greedy king game.
\newblock {\em Int. J. Mod. Phys. B}, 20:1390, 2006.

\bibitem{bargill2006ssm}
N.~Bar-Gill and G.~Kurizki.
\newblock {Signatures of Strong Momentum Localization via Entanglement of
  Translational and Internal States}.
\newblock {\em Physical Review Letters}, 97(23):230402, 2006.

\bibitem{TIE:OL}
Michal Kol\'{a}\v{r}, Tom\'{a}\v{s} Opatrn\'{y}, and Gershon Kurizki.
\newblock Path and phase determination for an interfering photon with orbital
  angular momentum.
\newblock {\em Optics Letters}, 33(1):67--69, 2008.

\bibitem{holevo2001ssq}
A.S. Holevo.
\newblock {\em {Statistical Structure of Quantum Theory}}.
\newblock Springer, 2001.

\bibitem{helstrom1969qda}
C.W. Helstrom.
\newblock {Quantum detection and estimation theory}.
\newblock {\em Journal of Statistical Physics}, 1(2):231--252, 1969.

\bibitem{peres1995qtc}
A.~Peres.
\newblock {\em {Quantum Theory: Concepts and Methods}}.
\newblock Kluwer Academic Pub, 1995.

\end{thebibliography}
\bibliographystyle{unsrt}

\end{document}